\documentstyle[12pt]{article}
\newcommand{\newsection}[1]{\section{#1}}

\def\al{\alpha}

\def\dl{\delta}                

\def\om{\omega}               
               
\def\Ac{\mbox{\protect$\cal A$}}
\def\Bc{\mbox{\protect$\cal B$}}
\def\Dc{\mbox{\protect$\cal D$}}
\def\Jc{\mbox{\protect$\cal J$}}

\def\Sc{\mbox{\protect$\cal S$}}

\def\Ab{{\bf A}}

\def\FF{I\!\!F}
\def\RR{I\!\!R}
\def\ZZ{Z\!\!\! Z}

\def\sp{{\rm sp}}

\def\goto{\rightarrow}

\def\wh#1{\widehat{(#1)}}
\def\inv{^{-1}}
\def\sds{\subset\hskip -1em +}  

\def\half{{\scriptstyle {1 \over 2}}}

\def\hs{\hspace{5mm}}
\def\hsc{\hspace{5mm},\hspace{5mm}}
\def\ie{{\em i.e.}}
\def\eg{{\em e.g.}}

\def\beq{\begin{equation}}
\def\eeq{\end{equation}}

\begin{document}


\begin{titlepage}

\begin{flushright}
RI-3-96\\
September, 1996\\[15mm]
\end{flushright}

\begin{center}
\LARGE
A New Family of Solvable Self-Dual Lie Algebras
\\[10mm]
\large
Oskar Pelc\\[5mm]
\normalsize
{\em Racah Institute of Physics, The Hebrew University\\
  Jerusalem, 91904, Israel}\\
E-mail: oskar@shum.cc.huji.ac.il
\\[15mm]
\end{center}

\begin{abstract}
A family of solvable self-dual Lie algebras is presented. 
There exist a few methods for the construction of non-reductive self-dual Lie
algebras: an orthogonal direct product, a double-extension of an Abelian 
algebra, and a Wigner contraction. It is shown that the presented algebras 
cannot be obtained by these methods. 
\end{abstract}

%

\end{titlepage}

\flushbottom


\newsection{Introduction}
A {\em self-dual Lie algebra} $\Ac$ is a finite-dimensional Lie algebra that 
admits an {\em invariant metric}, \ie, a symmetric non-degenerate bilinear 
form $(\cdot,\cdot)$ that
is invariant under the adjoint action of the corresponding group:
\beq (gx_1g\inv,gx_2g\inv)=(x_1,x_2), \hs \forall x_i\in\Ac, \eeq
for any $g$ in the group, or equivalently,
\beq\label{invariance}
  ([y,x_1],x_2)=-(x_1,[y,x_2]), \hs \forall x_i\in\Ac,
\eeq
for any $y\in\Ac$ (other common names for such algebras are ``orthogonal'' and 
``symmetric-self-dual'').

Self-dual Lie algebras are important in physics. In string theory,
one needs a (highest weight) representation of the Virasoro algebra 
\cite{Virasoro}. An important source of such representations is the 
affine Sugawara construction \cite{Sugawara}, starting from a Kac-Moody 
algebra \cite{Kac-Moody}. Self-dual Lie algebras are precisely the Lie 
algebras that are needed for such a construction \cite{Mohammedi,FOF-Stan}.
{}From the point of view of two-dimensional conformal quantum field theories, 
a self-dual Lie algebra is the mathematical structure needed for the 
construction of Wess-Zumino-Novikov-Witten (WZNW) models \cite{WZNW}.

The best known families of self-dual algebras are the semi-simple algebras 
[where the (essentially unique) invariant metric is the Killing form] and the 
Abelian algebras (for which every metric is trivially invariant). However, 
these are not the only ones. 
Any self-dual Lie algebra can be constructed, using semi-simple and Abelian 
algebras, by a sequence of construction steps, each of which is either an 
{\em orthogonal direct sum} (\ie,
a direct sum equipped with the natural direct 
sum metric) or a procedure called {\em double extension} \cite{Med-Rev}
(see also \cite{FOF-Stan}). 
Self-dual algebras that are not orthogonal direct sums are called
{\em indecomposable} and those that are also non-reductive and are not double 
extensions of an {\em Abelian} algebra will be called {\em deeper} (following
\cite{FOF-Stan}), since their construction involves more then one step of 
double-extension. Another method to obtain self-dual algebras is 
through a {\em Wigner contraction} \cite{Ino-Wig}, as described in \cite{ORS}. 
A self dual algebra obtained by this method is always a double extension of an
Abelian algebra and therefore is not a deeper algebra.

In this paper we present a new (as far as we know) infinite family of 
indecomposable, non-reductive (in fact, solvable), self-dual algebras 
$\{\Ac_{3m}\}$. We show that these algebras (except the first two)
are {\em deeper} algebras, in the sense defined above.
Among the known self dual algebras, deeper algebras are rare (in fact, we do 
not know any other examples). The reason for this may be the absence of a 
practical method to construct such algebras (as will be explained in section 
IV). Therefore the family presented here may provide a valuable test-ground 
in the study of the properties and structure of general self-dual algebras 
and the physical models based on them \cite{GPR}.

The algebra
\[ \Ac_n\equiv \sp\{T_i\}_{0\leq i\leq n} \]
is defined by the Lie bracket
\beq\label{lie}
  [T_i,T_j]=\left\{\begin{array}{cc}
    \wh{i-j}T_{i+j}	&	i+j\leq n, \\
    0			&	\mbox{otherwise},
  \end{array}\right.
\eeq
where $\hat{i}\equiv i \bmod 3$ is chosen to be in $\{-1,0,1\}$. 
When $\hat{n}=0$, the metric
\beq (T_i,T_j)=\dl_{i+j-n}+b\dl_i\dl_j \eeq
is an invariant metric on $\Ac_n$ (for arbitrary $b$).

In section II we define the algebras $\Ac_n$ and prove that (for $\hat{n}=0$) 
these are indeed self-dual Lie algebras. In section III we find all the ideals 
of $\Ac_n$. This result is used in section V, where we check which of these
algebras is a double extension of an Abelian algebra or a result of a 
Wigner contraction. Before that, in section IV, we describe these two methods
briefly, emphasizing the gap that the algebras presented here may help to
reduce. Finally, in section VI, we comment about possible generalizations.

\newsection{The Algebra $\Ac_n$}
\label{An}

Consider a vector space, equipped with the basis $\{T_i\}_{i\in\ZZ}$
and the following ``Lie bracket'':
\beq\label{Bracket-An} [T_i,T_j]=\wh{i-j}T_{i+j}, \eeq
where $\hat{i}\equiv i \bmod 3$ is chosen to be in $\{-1,0,1\}$. The map
$i\goto\hat{i}$ is {\em almost} a ring homomorphism $\ZZ\goto\ZZ$: it
preserves multiplication,
\beq\label{mult} \wh{ij}=\hat{i}\hat{j}, \eeq
and almost preserves addition:
\beq\label{add1} \wh{i+j}=\wh{\hat{i}+\hat{j}} \hsc \wh{-i}=-\hat{i}, \eeq
\beq\label{add2}  \wh{i-j}=0 \Longleftrightarrow \hat{i}=\hat{j} \eeq
(but note that $\hat{1}+\hat{1}\neq\wh{1+1}$ \cite{props}). These are the 
properties that will be used in the following. Particularly useful will be 
the property
\beq \hat{i}=\hat{j} \Longleftrightarrow \wh{i+k}=\wh{j+k}, \eeq
which follows from (\ref{add2}).

The bracket is manifestly anti-symmetric so to obtain a Lie algebra, there
remains to verify the Jacobi identity. Since
\beq
 [[T_i,T_j],T_k]=\hat{c}_{ijk}T_{i+j+k} \hsc
 c_{ijk}\equiv(i-j)(i+j-k),
\eeq
the Jacobi identity takes the form
\beq\label{jac} \hat{c}_{ijk}+\hat{c}_{jki}+\hat{c}_{kij}=0. \eeq
This identity holds without the `hats', therefore, by (\ref{add1}),
\beq\label{vir}  \hat{c}_{ijk}+\hat{c}_{jki}+\hat{c}_{kij}=0 \bmod 3, \eeq
so (\ref{jac}) can be false only when
\beq  \hat{c}_{ijk}=\hat{c}_{jki}=\hat{c}_{kij}=\pm1. \eeq
$\hat{c}_{ijk}=1$ is equivalent to $\wh{i-j}=\wh{i+j-k}=\pm1$ and,
therefore, also to
\[ i=j\pm1 \bmod 3 \hsc k=-j \bmod 3, \]
and this cannot hold simultaneously for all the tree cyclic permutations
of $\{ijk\}$. Replacing $i\leftrightarrow j$, one obtains the same
result for $\hat{c}_{ijk}=-1$. Therefore, the Jacobi identity holds and
the above algebra is indeed a Lie algebra (over the integers).

Let us consider the subalgebra
\[ \Ac_{\infty}\equiv \sp\{T_i\}_{i\geq0}. \]
Dividing by the ideal $\sp\{T_i\}_{i>n}$ (for some positive integer
$n$), one obtains the  finite-dimensional Lie algebra
\[ \Ac_n\equiv \sp\{T_i\}_{0\leq i\leq n} \]
with the Lie bracket
\beq
  [T_i,T_j]=\left\{\begin{array}{cc}
    \wh{i-j}T_{i+j}	&	i+j\leq n, \\
    0			&	\mbox{otherwise}.
  \end{array}\right.
\eeq
{}From now on we restrict our attention to such an algebra. It is a
solvable algebra, $T_0$ being the only non-nilpotent generator  and it
possesses a $\ZZ$-grading: deg$(T_i)=i$ (inherited from the original
infinite-dimensional algebra).

We would like to find an invariant metric $(\cdot,\cdot)$ on $\Ac_n$.
Using (\ref{lie}), the invariance condition
\[ ([T_k,T_i] , T_j) + (T_i , [T_k,T_j]) = 0 \]
takes the form
\beq\label{inv} \wh{k-i}(T_{k+i},T_j)+\wh{k-j}(T_{k+j},T_i)=0 \eeq
(here $T_i\equiv0$ for $i>0$) and, in particular, for $k=0$:
\beq \wh{\wh{-i}+\wh{-j}}(T_i,T_j)=0, \eeq
which, by eqs. (\ref{add1},\ref{add2}), is equivalent to
\beq\label{diag} \wh{i+j}(T_i,T_j)=0. \eeq
This means that two out of each three ``reversed''
(right-up-to-left-down) diagonals vanish. Let us look for a metric
with only one non-vanishing diagonal.
To obtain a non-degenerate form, this must be the central diagonal and
according to (\ref{diag}), this is possible only for $\hat{n}=0$. We,
therefore, concentrate on this case and consider a metric of the form
\beq (T_i,T_j)=\om_j\dl_{i+j,n}\hsc \om_{n-j}=\om_j\neq0. \eeq
For such a metric the invariance condition (\ref{inv}) takes the form
\beq \wh{k-i}\om_j+\wh{k-j}\om_i=0, \hs \forall i+j+k=n \eeq
and using $\hat{n}=0$, one obtains
\beq\label{om1} \wh{2i+j}\om_j+\wh{2j+i}\om_i=0. \eeq
First we take $\hat{j}=0$, which gives
\beq\label{met3} \hat{i}(\om_i+\hat{2}\om_j)=0, \eeq
and this implies (since $\hat{2}\neq0$)
\beq \label{om2}
  \om_i=\left\{\begin{array}{cc}
    \om_i=-\hat{2}\om_0	&      \hat{i}\neq0, \\
    \om_i=\om_0        	&      \hat{i}=0.
  \end{array}\right.
\eeq
Using this result we take $\hat{i},\hat{j}\neq0$ in (\ref{om1})
and obtain
\beq\label{met5} \hat{2}\cdot\hat{3}\wh{i+j}\om_0=0, \eeq
which is satisfied, since $\hat{3}=0$.
Also, since $-\hat{2}=1$, we have 
$\om_i=\om_0,\;\forall i$. To summarize, we proved the following.

\noindent{\bf Lemma:}
\begin{quote}\em
  A (non-degenerate) invariant metric on $\Ac_n$ with only one
  (reversed) diagonal exists iff $\hat{n}=0$ and it is proportional to
  \beq (T_i,T_j)=\dl_{i+j-n}. \eeq
\end{quote}
Note that one can add to the metric a multiple of the Killing form,
obtaining
\beq (T_i,T_j)=\dl_{i+j-n}+b\dl_i\dl_j \eeq
(with $b$ arbitrary). The appearance of the second term can also be
seen as a result of the (automorphic) change of basis
\[ T_0\goto T_0+\half bT_n. \]

\newsection{The Ideals in $\Ac_n$}
\label{Ideal}

In this section we continue to analyze the algebra $\Ac_n$,
looking for all its ideals and concluding that, for $\hat{n}=0$, the
only ideals are of the form
\[ \Ac_{m,n}\equiv \sp\{T_i\}_{i=m}^n. \]
This will be important later, when we will check if
these algebras are double extensions of Abelian algebras.
The grading on $\Ac_n$ (deg$(T_i)=i$) will play a central role in the
following and will be called ``charge''. The adjoint action of $T_i$
increases the charge by $i$. Note that there are only positive charges,
so that the adjoint action cannot decrease the charge. This
proves that $\Ac_{m,n}$ (for any $m$) is indeed an ideal.

Let $\Jc$ be an ideal in $\Ac_n$. We choose a basis for $\Jc$ such that 
each element has a {\em different} minimal charge (this can be easily 
accomplished) and, therefore, can be labeled by it. We, therefore, have
(after an appropriate normalization)
\beq
  \Jc=\sp\{S_{\al}\}, \hs S_{\al}-T_{\al}\in\Ac_{\al+1,n}.
\eeq
Isolating in $\Jc$ the {\em maximal} ideal of the form $\Ac_{m,n}$, we
obtain
\beq 
  \Jc=\sp\{S_{\al}\}_{\al\in\Ab}\bigoplus\Ac_{m,n} \hsc
  m-1\not\in\Ab.
\eeq
Observe that this implies that for any element in $\Jc$ that is not in
$\Ac_{m,n}$, its minimal charge is in $\Ab$.

The choice $\Ab=\emptyset$ (the empty set) corresponds to the ``trivial''
solution $\Jc=\Ac_{m,n}$. In the following we look for other solutions,
\ie, with $\Ab\neq\emptyset$. This also implies max$(\Ab)<m-1$.
We are going to explore the restrictions on the $S_{\al}$'s implied by
the claim that $\Jc$ is an ideal in $\Ac_n$. Since $\Ac_{m,n}$ is an ideal
by itself, the only restrictions come from
\beq [T_i,S_{\al}]\in\Jc \hs \forall\al\in\Ab\hsc i=0,\ldots,n. \eeq
$\Jc$ contains all terms with charge of at least $m$, therefore, restrictions
will arise only in terms in the commutator with smaller charge. For
$i\geq m-\al$ there are no such terms. As the charge $i$ decreases,
there will be more non-trivial terms, therefore, we will start from the
higher charges.

For $i=m-\al-1$ we have (in the following, ``$\simeq$'' means ``equality
up to an element of $\Ac_{m,n}$''):
\beq
  [T_{m-\al-1},S_{\al}]\simeq
  [T_{m-\al-1},T_{\al}]=\wh{m-2\al-1}T_{m-1}
\eeq
(here and in other similar cases the hat should be applied to the
whole expression between parentheses).
$T_{m-1}\not\in\Jc$ (otherwise $\Ac_{m-1,n}\subset\Jc$), therefore,
\beq \wh{m-2\al-1}=0. \eeq
Using eqs. (\ref{add1},\ref{add2}), this is equivalent to
\beq\label{halbt} \hat{\al}=-\wh{2\al}=-\wh{m-1}, \eeq
and since this is true for all $\al\in\Ab$, we also have
\beq\label{halal}
  \hat{\al}_1=\hat{\al}_2, \hs \forall\al_1,\al_2\in\Ab.
\eeq

Next, for $i=m-\al-2$ we have [using eqs. (\ref{halbt}) and
(\ref{add1})]
\beq\label{second}
  [T_{m-\al-2},S_{\al}]\simeq
  [T_{m-\al-2},T_{\al}+s_{\al}^{\al+1}T_{\al+1}]=
  -T_{m-2}+s_{\al}^{\al+1}T_{m-1}.
\eeq
This implies that $m-2$ is a minimal charge of an element of $\Jc$,
therefore, $m-2\in\Ab$. Substituting $\al=m-2$ in (\ref{second}), we
obtain
\beq
  [T_0,S_{m-2}]\simeq
  -T_{m-2}+s_{m-2}^{m-1}T_{m-1}\simeq
  -S_{m-2}+2s_{m-2}^{m-1}T_{m-1},
\eeq
and this implies $s_{m-2}^{m-1}=0$, so with no loss of generality, we
can choose
\beq S_{m-2}=T_{m-2}. \eeq

Finally, for $i=m-\al-3$ and $m-2>\al\in\Ab$, we have
\beq
  [T_{m-\al-3},S_{\al}]\simeq
  [T_{m-\al-3},T_{\al}+s_{\al}^{\al+1}T_{\al+1}+s_{\al}^{\al+2}T_{\al+2}]=
  T_{m-3}+s_{\al}^{\al+2}T_{m-1},
\eeq
and as before this should imply that $m-3\in\Ab$ (being the minimal
charge of an element of $\Jc$). However, according to eq. (\ref{halal}),
this is impossible since $m-2\in\Ab$. Therefore, $\Ab$ contains no
elements other then $m-2$ and $\Jc$ is of the form
\beq \Jc=\sp\{T_{m-2}\}\oplus\Ac_{m,n}. \eeq
A straightforward check [or use of eq.\ (\ref{halbt})] shows that this
is indeed an ideal iff $\hat{m}=0$. 

Is this ideal really non-trivial? It turns out that, for $\hat{n}\neq 1$
(including the case of main interest to us: $\hat{n}=0$), it is not! 
To see this, consider the (non-singular) linear map defined by 
$T_i\mapsto T'_i\equiv-T_{i+\hat{i}}$ (note that for $\hat{n}=1$ this is not
well defined). Since $\hat{m}=0$, this map transforms 
$\Jc$ to $\Ac_{m-1,n}$
\beq
  [T'_i,T'_j]=-\wh{i-j}T_{i+j+(\hat{i}+\hat{j})}
             =-\wh{i-j}T_{i+j+\wh{i+j}}=\wh{i-j}T'_{i+j}
\eeq
[the second equality follows from the fact that for $\wh{i-j}\neq0$,
$\wh{i+j}=\hat{i}+\hat{j}$], therefore, this map is an automorphism of
Lie algebras, which means that
$\Jc=\sp\{T_{m-2}\}\oplus\Ac_{m,n}$ is automorphic to $\Ac_{m-1,n}$.

\newsection{Construction Methods of Self-Dual Lie Algebras}

Our next goal is to show how the self-dual algebras described above (\ie, with
$\hat{n}=0$) fit into the general family of self-dual algebras, and to
clarify their significance. 
As noted in the introduction, it has been proved \cite{Med-Rev} that 
any self-dual Lie algebra can be constructed, using semi-simple and Abelian 
algebras, by a sequence of construction steps, each of which is either an 
{\em orthogonal direct sum} (\ie,
a direct sum equipped with the natural direct 
sum metric) or a procedure called {\em double extension} \cite{Med-Rev}. This
result seems, at first sight, to make all self-dual algebras available to us,
but practically, this is not so, as we now explain.

The {\em double extension of a self-dual Lie algebra $\Ac$ by another
Lie algebra $\Bc$} (not necessarily self-dual) can be seen as a two-step
process (we will give here only the information that will be needed later; 
more details can be found, for example, in \cite{GPR}). The first step is to 
combine $\Ac$ and $\Bc$ to a {\em semi-direct sum}
\beq \Sc=\Bc\sds\Ac \eeq
(\ie, $\Sc$ is a direct sum of the vector spaces $\Bc$ and $\Ac$, $\Bc$ is 
a subalgebra, and $\Ac$ is an ideal) in such a way that the metric 
in $\Ac$ will be invariant also under the action of $\Bc$. 
The second step is the extension of $\Sc$ by an Abelian algebra
$\Bc^*$ with $\dim\Bc^*=\dim\Bc$. The resulting algebra $\Dc$ has a Lie 
product of the following general form:
\beq \begin{array}{c|ccc}
  [\cdot,\cdot] &  \Bc   &  \Ac       &  \Bc^* \\ \hline
  \Bc           &  \Bc   &  \Ac       &  \Bc^* \\
  \Ac           &  \Ac   &  \Ac+\Bc^* &  0  \\
  \Bc^*         &  \Bc^* &  0         &  0
\end{array}\eeq
For the first step one needs, in addition to the algebras $\Ac$ and $\Bc$, a
representation of $\Bc$ as derivations in $\Ac$. Moreover, it was shown in 
\cite{FOF-Stan} that if $\Bc$ acts through {\em inner} derivations (\ie,
the action of each $y\in\Bc$ coincides with the adjoint action of an
element $\hat{y}\in\Ac$: $y:x\goto[\hat{y},x]$), the resulting algebra $\Dc$
is decomposable (\ie, expressible as an orthogonal direct sum). This means that
for the construction of an indecomposable
double extension, one needs knowledge about the {\em outer}
(non-inner) derivations in $\Ac$, and such information is not available
in general. Therefore a discovery of unknown (indecomposable) 
self-dual algebras is indeed significant.

Another method for constructing new self-dual Lie algebras is by
performing a {\em Wigner contraction} \cite{Ino-Wig} (this was proposed, in 
the context of string theory, in \cite{ORS}).
The initial data for this construction consists of a self-dual Lie
algebra $\Sc_0$ and a subalgebra $\Bc_0$ of $\Sc_0$ such that the
restriction
of the metric on $\Sc_0$ to $\Bc_0$ is non-degenerate.
Unlike in double extensions, the initial data needed here is very simple and 
generally available (for example, one can take $\Sc_0$ and $\Bc_0$ to be 
simple), therefore, the method can be easily used to find many 
new non-trivial self-dual algebras. It turns out, however \cite{FOF-Stan}, 
that the resulting algebra is always a double extension of an
{\em Abelian} algebra. Actually, when $\Ac$ (in the process of double 
extension) is Abelian, the problem of the initial data described above does 
not exist \cite{Abel} and one can indeed construct large families of double 
extensions of such algebras. Therefore the non-trivial task is to find 
(indecomposable, self-dual) {\em deeper} algebras (as they were called in 
\cite{FOF-Stan}): algebras that their 
construction out of simple and one-dimensional algebras involves
more than one double extension \cite{Reduct}. In the next section we show 
that almost all the algebras defined in section \ref{An} are such algebras.

\newsection{$\Ac_n$ as a Deeper algebra}
In the previous section we described the following inclusion relations among 
the indecomposable self-dual algebras:
\begin{center}
  $\{$ Indecomposable, Self-Dual Algebras $\}$ \\
  $\cup$ \\
  $\{$ (Single) Double-Extensions of Abelian Algebras $\}$ \\
  $\cup$ \\
  $\{$ Algebras obtainable by a Wigner contraction $\}$.
\end{center}
The results of this section will imply that these are {\em strict} inclusions,
\ie, all the three sets are distinct. Explicitly we will show here that among 
the algebras $\Ac_n$ with $\hat{n}=0$ (which were shown in section \ref{An} to
be self-dual), $\Ac_3$ can be
obtained by a Wigner contraction, $\Ac_6$ is a double extension of an
Abelian algebra but {\em cannot} be obtained by a Wigner contraction, and the
rest are deeper algebras, \ie\ they are {\em not} double extensions of
Abelian algebras and, therefore, in particular, they cannot be
obtained by a Wigner contraction.

First, from the list of the ideals found in section \ref{Ideal} we observe 
that $\Ac_n$ is indeed indecomposable \cite{Double}. 
Next, we try to identify in $\Ac_n$ the structure of a double
extension of an Abelian algebra. The Lie product in an algebra $\Dc$
with such a structure is of the form
\beq \begin{array}{c|ccc}
  [\cdot,\cdot] &  \Bc   &  \Ac   &  \Bc^* \\ \hline
  \Bc           &  \Bc   &  \Ac   &  \Bc^* \\
  \Ac           &  \Ac   &  \Bc^* &  0  \\
  \Bc^*         &  \Bc^* &  0     &  0
\end{array}\eeq
In this table we recognize two properties of $\Dc$:
\begin{enumerate}
  \item $\Dc$ is a semi-direct sum of $\Bc$ and the ideal
    $\Jc=\Ac+\Bc^*$: $\Dc=\Bc\sds\Jc$;
  \item $[\Jc,\Jc]\subset\Bc^*$, therefore,
    $\dim[\Jc,\Jc]\leq\dim\Bc^*=\dim\Bc$.
\end{enumerate}

Consider the first property. The candidates for the ideal
$\Jc$ were found in the previous section. It was shown that
$\Jc=\Ac_{m,n}$ (possibly after an automorphic change of basis
$\{T_i\}$). Following the same approach, we choose a basis
$\{R_i\}_{i=0}^{m-1}$ for $\Bc$ such that $i$ is the minimal charge
of $R_i$. $[T_{m-1},T_{m-2}]=T_{2m-3}$ and $2m-3<n$ (since
dim$\Ac_n\geq2$dim$\Bc$), therefore, $[R_{m-1},R_{m-2}]\neq0$ and its
minimal charge is $2m-3$. $\Bc$ is
closed under the Lie bracket and $\Bc\cap\Jc=\{0\}$, therefore,
$[R_{m-1},R_{m-2}]\not\in\Jc$, which implies that $2m-3<m$.
This leaves us with $m=1$ or $2$ \cite{m-null}.

As for the second property, we have
\beq\label{JJm} dim[\Jc,\Jc]\leq\dim\Bc=m. \eeq
One can easily verify that
\beq [\Ac_{m,n},\Ac_{m,n}]=\Ac_{2m+1,n}, \eeq
therefore, eq. (\ref{JJm}) implies $n\leq3m$. On the other hand,
$n+1\geq2m$ (since $\dim\Ac_n\geq2\dim\Bc$). Recalling that
$\hat{n}=0$, We obtain three possibilities:
\beq\label{pos} (m,n)=(1,3),(2,3),(2,6), \eeq
and a direct check confirms that each of them indeed corresponds to a
double extension of an Abelian algebra $\Ac$ (in the second possibility
$\Ac$ is zero-dimensional). Observe that there is more
than one way to represent an algebra as a double extension. Moreover,
$\Ac_6$ can be obtained both by extending an Abelian algebra (with
$m=2$) and by extending a non-Abelian algebra (with $m=1$), so the
number of double extensions leading to a given Lie algebra is not
unique \cite{Depth}.

Turning to the search of the structure of a Wigner contraction, the only
candidates are those enumerated in (\ref{pos}). $\Ac_3$ is the Heisenberg 
algebra, and it is indeed a Wigner contraction of $so(2,1)\oplus so(2)$ 
[which leads to the first possibility in (\ref{pos})]. The other candidate 
is $\Ac_6$, which corresponds to the last possibility in (\ref{pos}).
To examine this case, we use the further requirement that in a Wigner 
contraction, $\Bc$ must be self-dual \cite{BCD}.
For $m=2$, $\Bc$ is the two-dimensional, non-Abelian Lie algebra,
\[ [R_0,R_1]=R_1. \]
This algebra is not self-dual, therefore, even if $\Ac_6$ can be obtained 
by a Wigner contraction, this procedure will not lead to an invariant 
metric on $\Ac_6$. 

\newsection{Generalizations of the algebras $\Ac_n$}

We conclude with some comments about possible generalizations of the algebras 
defined in section \ref{An}, obtained by using the defining relations
(\ref{Bracket-An}) with a {\em different} choice of the map
``$\hat{\hspace{3mm}}$''. If one takes ``$\hat{\hspace{3mm}}$'' to
be some homomorphism from $\ZZ$ to some commutative ring $\FF$ with
unity, (\ref{mult}-\ref{add2}) hold, as well as (\ref{jac}) and one
obtains a Lie algebra over $\FF$. For example, one can take $\FF=\ZZ_p$
($p$ a positive integer) with ``$\hat{\hspace{3mm}}$'' being the
natural homomorphism \cite{Phys}.
Another example is obtained by taking $\FF=\ZZ$ and ``$\hat{\hspace{3mm}}$'' 
the identity map, the result being the Virasoro algebra (with zero central 
charge). A different approach would be to keep $\FF=\ZZ$ and to look for a map 
``$\hat{\hspace{3mm}}$'' (not necessarily a homomorphism) with the required 
properties. A natural candidate would be $\hat{i}=i \bmod p$ ($p$ a positive 
integer). Taking $p=2$ and $\hat{\hspace{3mm}}:\ZZ\goto\{0,1\}$, an analysis 
similar to the $p=3$ case leads to the choice $(i,j,k)=(1,0,0)$, for which the
right-hand side of (\ref{jac}) does not vanish
($\hat{c}_{ijk}=\hat{c}_{kij}=1$, $\hat{c}_{jki}=0$). There seems to
be no other choice of $p$ and range of the map ``$\hat{\hspace{3mm}}$''
such that the multiplication is preserved. 

In the previous sections we referred to the specific choice 
$\hat{i}=i \bmod 3\in\{-1,0,1\}$,
but actually section \ref{An} can be easily extended to a general commutative 
ring $\FF$ with unity that has no zero-divisors and a 
general map ``$\hat{\hspace{3mm}}$'' satisfying properties 
(\ref{mult}-\ref{add2}) and the Jacobi identity (\ref{jac}). The result is
that the statement of the lemma is true whenever $\hat{3}=0$, while for
$\hat{3}\neq0$ (and, in particular, for $\hat{i}=i$) there is no invariant 
metric with only one diagonal [for $\hat{2}=0$, (\ref{met3}) fails while for 
$\hat{2},\hat{3}\neq0$, (\ref{met5}) fails]. 
In most of section \ref{Ideal} the only additional assumption used is 
$\hat{3}=0$. Only the automorphism described at the end of the section assumes
the explicit form of ``$\hat{\hspace{3mm}}$''.

\vspace{1cm}
\noindent{\bf Acknowledgments}

We would like to thank A. Reznikov and J. Bernstein for helpful discussions.


\begin{thebibliography}{99}
\bibitem{props}
   The $\Longleftarrow$ direction of (\ref{add2}) follows from
   (\ref{add1}), but for the other direction (\ref{add1}) only implies
   \[ \wh{i-j}=0 \Longrightarrow \hat{i}-\hat{j}=0 \bmod 3, \]
   and the stronger result $\hat{i}-\hat{j}=0$ follows from the fact
   that $|\hat{i}-\hat{j}|$ is always at most $2$. When $\wh{i-j}\neq0$
   this reasoning breaks down and indeed we have, \eg,
   $\hat{2}-\hat{1}\neq\wh{2-1}$.
\bibitem{Abel}
   For Abelian $\Ac$, any (non-trivial) linear map in $\Ac$ is obviously an 
   outer derivation, and to construct a double extension one only 
   needs to worry about the invariance of the metric (the second step of the 
   construction is completely determined by the first
   one and poses no further difficulties).
\bibitem{Double} 
   This means \cite{Med-Rev} that they should be expressible 
   as double extensions by the one-dimensional algebra, and this structure 
   can be indeed easily identified with:
   \[ \Bc=\sp\{T_0\} \hsc \Bc^*=\sp\{T_n\} \hsc \Ac=\Ac_{1,n}/\Bc^*. \]
\bibitem{m-null}
   The value $m=0$ is also a possibility, but it is not interesting. 
   It corresponds to $\dim\Bc=0$. 
   As a double extension it means not to do anything --
   remaining with the (Abelian) algebra $\Ac$ one started with. As a
   Wigner contraction it means that, starting with some self-dual Lie
   algebra $\Sc_0$, all we did is to set its Lie bracket to $0$, so
   that we end up with the Abelian Lie algebra of the same dimension,
   which is trivially self-dual. In the present context, this
   corresponds to the one-dimensional algebra: $\Dc=\Ac_0$.
\bibitem{Phys} 
   This example, however, is irrelevant for string-theoretical models, since 
   there one needs an algebra over $\RR$ and for this $\FF$ must be some 
   sub-ring of $\RR$.
\end{thebibliography}
\end{document}